\documentclass[12pt]{article}
\usepackage{epsfig}

\def\beq{\begin{equation}}
\def\eeq#1{\label{#1}\end{equation}}
\def\eeqn{\end{equation}}

\def\beqa{\begin{eqnarray}}
\def\eeqa#1{\label{#1}\end{eqnarray}}
\def\eeqan{\end{eqnarray}}

\let\bar=\overbar

\def\Dslash{\not{\hbox{\kern-4pt $D$}}}
\def\dslash{\not{\hbox{\kern-2pt $\del$}}}

\def\msb{{\bar{\ssstyle M \kern -1pt S}}}

\def\Title#1{\begin{center} {\Large {\bf #1} } \end{center}}

\begin{document}

\Title{$K\rightarrow 3\pi$ decay results by NA48/2 at CERN SPS}

\begin{center}{\large \bf Contribution to the proceedings of HQL06,\\
Munich, October 16th-20th 2006}\end{center}

\bigskip\bigskip


\begin{raggedright}  

{\it Gianluca Lamanna\index{Lamanna, G.}\\
Department of Physics and INFN\\
University of Pisa\\
I-56127 PISA, ITALY}
\bigskip\bigskip
\end{raggedright}

\section{Introduction}
CP violation plays an important role in particle physics since its discovery 40 years ago~\cite{christ}. 
For more than 20 years this phenomenon appeared as confined in a particular sector of particle physics, through
the mixing between states of opposite CP in the neutral kaons.
The unambiguous discovery in the late 1990s, after the early indication by NA31~\cite{na31}, of direct CP violation in 
$K\rightarrow 2\pi$ decay, by the NA48~\cite{na48} and KTEV~\cite{ktev} experiments and the discovery of CP violation, in its various forms, 
in the neutral B meson system \cite{bdir} represented important breakthrough in the understanding of the particles dynamics.
A complete as possible study of the tiny effects due to the violation of this symmetry in all the systems where it can be carried out, represents
an important window on the contribution of new physics beyond the Standard Model: in fact new effects could appear, in particular in the 
heavy quark loops which are the core of the mechanism allowing the CP violation in the mesons decay. In the
kaon sector the most promising places, besides $\varepsilon^{'}/\varepsilon$, where this kind of contributions could play some role are the rates
of GIM suppressed rare decays and the charge asymmetry between charged kaons. In particular the $K\rightarrow 3 \pi$ asymmetry could give a 
strong qualitative indication of the validity of the CKM description of the direct CP violation or reveal the existence of possible 
sources outside this paradigm.

In principle Direct CP violation in $K^{\pm}$ can be detected comparing the different decay amplitudes in the $3\pi$ mode:
$$
|A(K^+\rightarrow\pi^+\pi^+\pi^-)|\neq|A(K^-\rightarrow\pi^-\pi^+\pi^-)|
$$
Experimentally the easiest way to study the difference between the charge conjugate modes is to compare the shape
of the \emph{Dalitz Plot} distribution instead of the decay rates. The small phase space in the 
three pion decay mode allows to expand the matrix element in terms of the, so called, \emph{Dalitz variables} u and v:
$$
u=\frac{(s_3-s_0)}{m^2_{\pi}} \qquad v=\frac{(s_2-s_1)}{m^2_{\pi}}
$$
These variables are defined using the Lorentz invariant $s_i=(p_K-p_i)^2$, where $p_K$ is the kaon four momentum and $p_i$ are 
the pion four momenta ($i=1,2,3$ the latter being ``odd'' pion) and $s_0=(s_1+s_2+s_3)/3$. Exploiting the Dalitz variables the
matrix element can be written as:
\begin{equation}\label{eqn:matrixel}
|M(u,v)|^2\sim 1 + gu + hu^2+kv^2+...
\end{equation}
where g,h,k are the linear and quadratic slope parameters. Using the fact that $|g|>>|h|,|k|$, the CP violation parameter
\begin{equation}\label{eqn:ag}
  A_g=\frac{g^+-g-}{g^++g^-}
\end{equation}
is defined using only the linear slopes given in (\ref{eqn:matrixel}). Being $g^+$ relative to $K^+$ decay while
$g^-$ to $K^-$, the parameter defined above is different from 0 only if an asymmetry exists between the matrix element 
describing kaon decay of opposite charge.
Theoretical predictions for the $A_g$ parameter both in the $K\rightarrow\pi^{\pm}\pi^0\pi^0$ ($A_g^n$, the so called ``neutral'' mode)
and in the $K\rightarrow\pi^{\pm}\pi^+\pi^-$ ($A_g^c$, the so called ``charged'' mode), are very difficult and the available predictions varying from $10^{-6}$ 
to few $10^{-5}$ are unreliable; calculations in the framework of theories beyond the standard model predict
a substantial enhancement of this parameter up to the level of few $10^{-4}$. Several experiments in the past have searched for asymmetry both 
in ``charged'' and ``neutral'' mode. The sensitivity reached so far is at level of $10^{-3}$, as summarized in table~\ref{tab:exper}.
\begin{table}[!t]
\begin{center}
\begin{tabular}{|l|l|l|}
\hline
\emph{Asymmetry} & \emph{\# of events} & \emph{Experiment} \\
\hline
$A_g^0=(19\pm125)\cdot10^{-4}$ & 115K & CERN PS(1975)~\cite{Smith:1973bi}\\
$A_g^0=(2\pm19)\cdot10^{-4}$ & 620K & Protvino IHEP (2005)~\cite{Akopdzhanov:2004xb} \\
\hline
$A_g^c=(-70\pm53)\cdot10^{-4}$ & 3.2M & BNL AGS (1970)~\cite{Ford:1970ta} \\
$A_g^c=(22\pm15\pm37)\cdot10^{-4}$ & 54M & HyperCP (2000) prelim~\cite{hypcp}\\
\hline
\end{tabular}
\end{center}
\caption{Summary of the experimental situation both in ``neutral'' ($A_g^0$) and in ``charged'' ($A_g^c$) mode, before the NA48/2 results}
\label{tab:exper}
\end{table}
The main goals of the NA48/2 experiment are to reach the sensitivity of $10^{-4}$ both in ``neutral'' and in ``charged'' mode, to investigate
the possibility of non standard model contributions to the CP violation in charged kaon decays, thus covering the gap existing between 
the experimental results and the SM theoretical predictions.
The large amount of $K\rightarrow 3\pi$ collected by NA48/2 experiment, allows a very precise measurement of the Dalitz plot
parameters  and, as recently shown~\cite{Cabibbo:2004gq}, the study of the neutral Dalitz plot density allows to extract important informations about the
pion scattering length. 

\section{Beams and detectors}
\begin{figure}[!tb]
\begin{center}
\epsfig{file=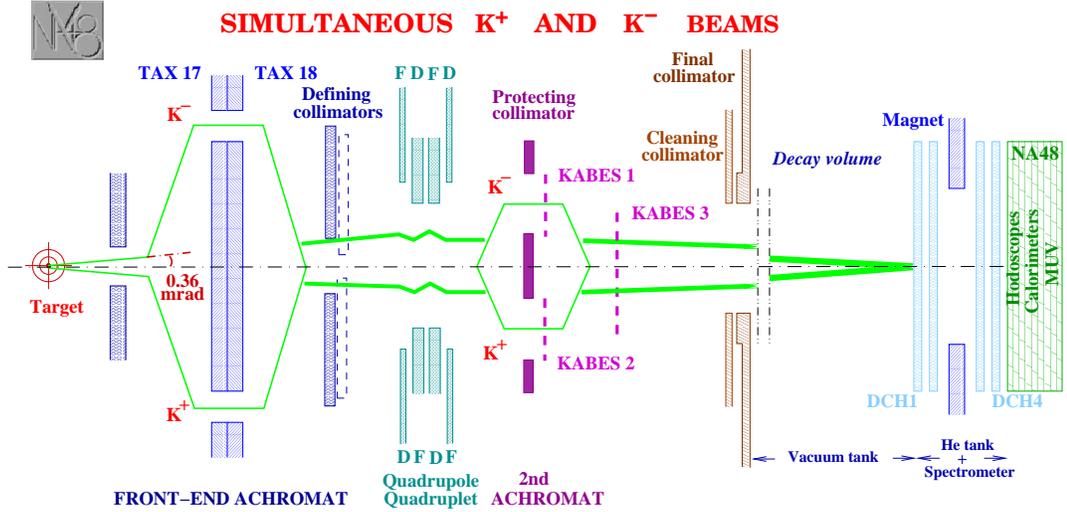,height=2.7in}
\caption{The NA48/2 beam line and detector. Not in scale.}
\label{fig:beam}
\end{center}
\end{figure}
The simultaneous collection of $K^+$ and $K^-$ decay with the same apparatus is an essential point of the asymmetry measurement.
A novel high intensity beam line was designed in the high energy hall (ECN3) at CERN SPS (fig.~\ref{fig:beam}). The charged particles (kaons, pions, muons
and electrons) are produced by 400 GeV high intensity protons beam($\sim 7\cdot10^{11}$ protons per pulse), 
from the SPS accelerator, with a 40 cm long and 2 mm in diameter beryllium target with an angle of zero degrees. A magnetic device, called 
\emph{first achromat}, selects the momentum of the beam in the range $(60\pm3) GeV/c$, splitting the two charges. After being recombined 
the beams are focused by a quadruplet of quadruples, before another splitting in the \emph{second achromat}. The second achromat houses
the first detector along the beam line, called KABES~\cite{peyo}, a spectrometer measuring the particles' momentum with a resolution of $\sim1\%$ 
(this detector is used studying rare decays). The two charged beams, recombined again along the beam axis, contain $\sim 6.4\cdot10^7$ particles
per 4.8 s spill, with a charge ratio $K^+/K^- \sim 1.8$ (irrelevant 
for the charge asymmetry measurement) and 
12 times more pions with respect to the kaons: however the pions decay products remain into the beam pipe, because of the small momentum and do not cross 
the detectors.
The decay region is housed in an evacuated tube $\sim 114 m$ long and $\sim2m$ in diameter. The beams in the 
decay region are superimposed within 1 mm with a total width of $\sim 5 mm$, so that both $K^+$ and $K^-$ decays illuminate with the 
same acceptance the same detector. The central detector is based on the old NA48 detector, described elsewhere~\cite{na48det}. For the 
asymmetry and Dalitz plot parameters measurement the two main detector are the spectrometer and the LKr calorimeter. The magnetic spectrometer
works with a $p_T=120MeV/c$ kick and the resolution in momentum (GeV/c) is $\sigma_p/p=1.0\%+0.044\%$. In order to manage the higher
intensity with respect to the previous NA48 runs, the drift chambers read out has been rebuilt. In order to collect the gammas coming from the
neutral pions decay a Liquid Krypton (LKr) calorimeter with a resolution in energy (GeV) of
$$
\frac{\sigma_E}{E}=\frac{3.2\%}{\sqrt E}+\frac{9\%}{E}+0.42\%
$$
is employed.
The very good resolution in the reconstructed kaon mass (1.7 $MeV/c^2$ for $K^{\pm}\rightarrow\pi^{\pm}\pi^+\pi^-$ and 0.9 $MeV/c^2$ for
$K^{\pm}\rightarrow\pi^{\pm}\pi^0\pi^0$) allows a precise calibration and monitoring of the characteristics and performances 
of the apparatus.
The data collection is based on a multilevel trigger system. The first level (L1) uses the information coming from a plastic scintillator
hodoscope and from a dedicated LKr readout, in which the number of peaks in the calorimeter's energy deposit are computed. The second level (L2) is based on processors
for a fast DCH reconstruction. In particular the number of reconstructed vertexes with 2 or 3 tracks, is used to  collect
 $K^{\pm}\rightarrow\pi^{\pm}\pi^+\pi^-$ events and the missing one-track mass, assuming the nominal kaon momentum (60 GeV) and direction (z axis), is 
used to collect $K^{\pm}\rightarrow\pi^{\pm}\pi^0\pi^0$, rejecting the main $K^{\pm}\rightarrow\pi^{\pm}\pi^0$ background.
NA48/2 collected data during two runs in 2003 and 2004, for a total of $\sim100 $ days of data taking. About $18\cdot10^9$ triggers
have been registered on tape, for a total of more than 200 TB.
 
\section{Charge asymmetry measurement strategy}
The asymmetry method is based on the comparison between the u projection of the Dalitz plot distribution, in order to extract the
difference between the matrix element linear components. This difference, defined as $\Delta g=g^+-g^-$, can be extracted
considering the ratio between the density in the u distribution for $K^+$ and $K^-$ decays. The ratio between the two distribution
can be written as:
\begin{equation}\label{ratio1}
R(u)=\frac{N_{K^+}}{N_{K^-}}\propto \frac{1+(g+\Delta g)u+hu^2}{1+gu+hu^2}
\end{equation}
where $g=(g^++g^-)/2$. From $\Delta g$ the asymmetry parameter $A_g$ can be easily evaluated using the relation $A_g\sim\Delta g/2 g$.
The simultaneous collection of decays coming from beams with similar momentum spectrum and a similar detection efficiency and acceptance
of the decay products, are fundamental points to control the instrumental charge asymmetry. However, the presence of magnetic fields
both in the beam sector (achromat) and in the detector (spectrometer magnet) could introduce an intrinsic charge dependent 
acceptance of the apparatus. To equalize this asymmetry the main magnetic fields were frequently reversed during the data taking.
During the 2003 run the magnet spectrometer polarity was reversed every day while the achromat magnets polarities every week. In the 2004
run the reversal was more frequent: every about 3 hours for the analyzing magnet and 1 day for the beam transport line magnets.
It is possible to redefine the ratio in (\ref{ratio1}) taking into account the magnetic field alternation. For instance, for a given
achromat polarities, the ratios:
\begin{equation}\label{ratio2}
R_J(u)=\frac{N^{B^-}_{K^+}}{N^{B^+}_{K^-}} \mbox{   ,} \qquad  R_S(u)=\frac{N^{B^+}_{K^+}}{N^{B^-}_{K^-}}
\end{equation}
are defined using the same side of the spectrometer, in the sense that the numerator and the denominator in these ratios
contain particles deflected in the same direction. The subscripts J and S represent the particles bending direction according to the 
geographic position of the Jura (J) and Saleve (S) mountains, respectively on the left and right side, with respect to the NA48/2 beam line direction.
Considering the possible achromat polarity, four independent ratios can be built exploiting the four different 
field combinations: instead of the single ratio (\ref{ratio1}), a \emph{quadruple ratio} can be defined as:
\begin{equation}\label{ratio4}
R(u)=R_{US}R_{UJ}R_{DS}R_{DJ}\sim \bar{R}\left( \frac{1+(g+\Delta g)u+hu^2}{1+gu+hu^2}\right) ^4
\end{equation}
where U and D stand for the path, up or down,  followed by $K^+$ in the achromat system, and $\bar{R}$ is an inessential normalization
constant. This method is independent on the relative size of the four samples collected with different fields configuration and on the
$K^+$ and $K^-$ flux difference.
In the value of $\Delta g$ extracted from the quadruple ratio (\ref{ratio4}), the benefits due to the polarity reversal 
are fully exploited and the main systematic biases due to instrumental asymmetries cancel out. In particular in the quadruple ratio 
there is a three fold cancellation:
\begin{itemize}
\item local detector bias (left-right asymmetry), thanks to the fact that each single ratio is defined in the same 
side of the detector;
\item beam local biases, because in each single ratio the path followed by the particle 
through the achromat is the same;
\item global time variation, because the decays from both charges are collected at the same time (this is not true for the same 
single ratio in which the numerator and the denominator are collected in different period).
\end{itemize}
The result remains sensitive only to the time variation of the detector with a characteristic time smaller than the inversion 
period of the magnetic field, if this effect is charge asymmetric and u dependent. 
Other systematics biases induced by effects not canceled by the magnetic field alternation of the magnetic fields (for instance 
the Earth's magnetic field in the decay region and any misalignment of the spectrometer) have been carefully corrected. The intrinsic
cancellation of instrumental asymmetries in the quadruple ratio allows to avoid the use of a MonteCarlo simulation. Nevertheless a GEANT3 
based MonteCarlo, including the full geometry description and time variations of beam characteristics, DCH inefficiency and spectrometer
alignment, is used for systematics studies and as a cross-check of the result.
   
\subsection{Results in  $K^{\pm}\rightarrow\pi^{\pm}\pi^0\pi^0$}
\begin{figure}[!thb]
\begin{minipage}[t]{0.46\textwidth}
\centering
\includegraphics[width=\textwidth]{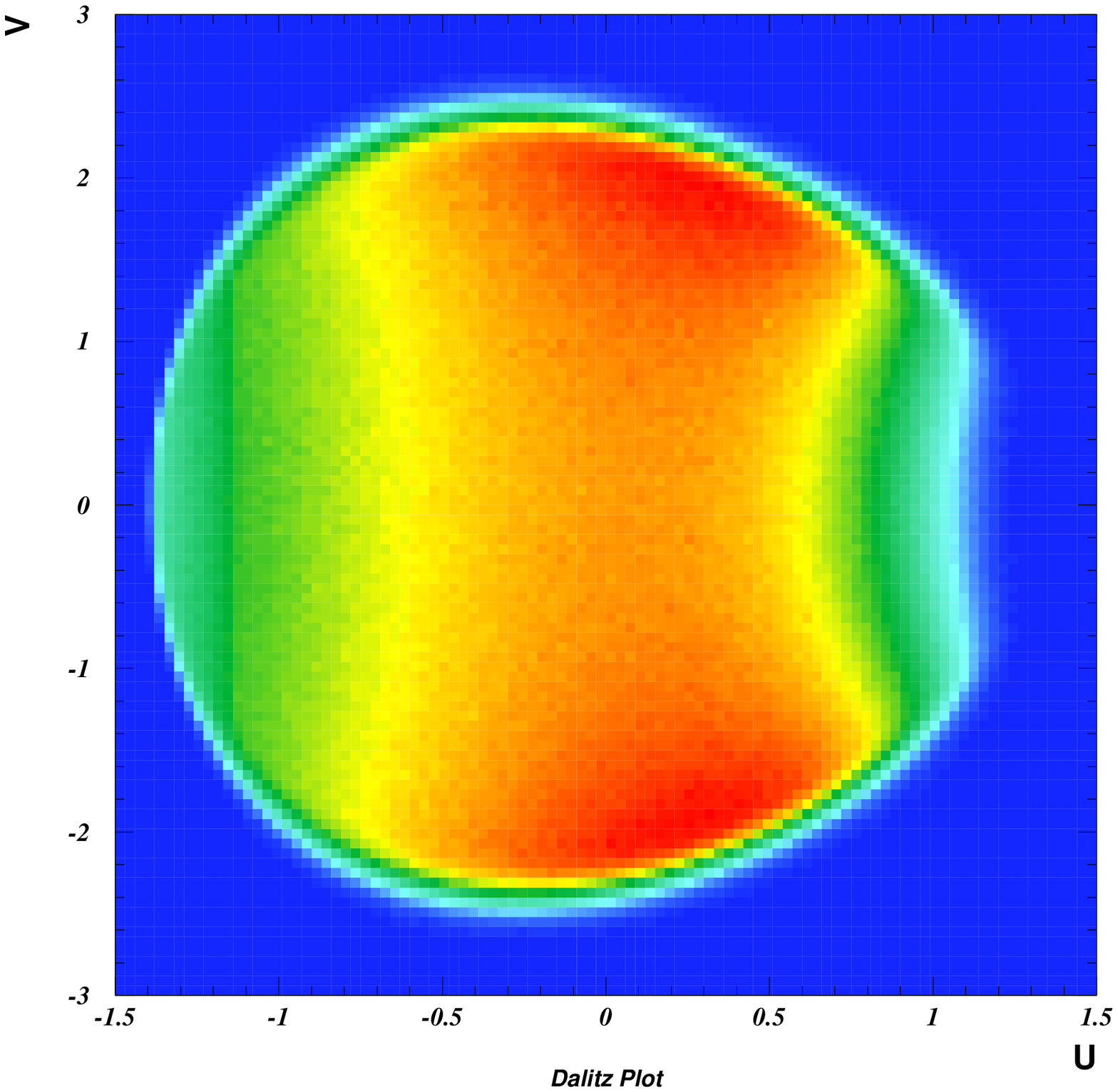}
\caption{\small  Dalitz plot distribution for  $K^{\pm}\rightarrow\pi^{\pm}\pi^0\pi^0$}
\label{fig:dalpn}
\end{minipage}\hfill
\begin{minipage}[t]{0.46\textwidth}
\centering
\includegraphics[width=\textwidth]{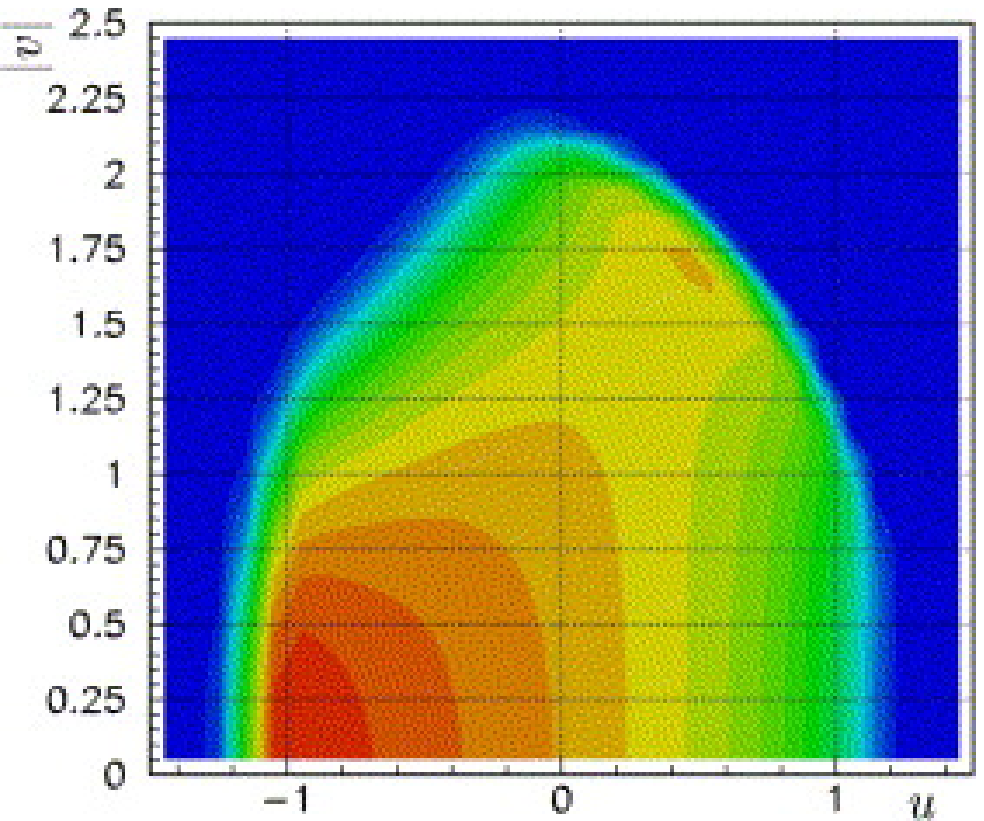}
\caption{\small Dalitz plot distribution for  $K^{\pm}\rightarrow\pi^{\pm}\pi^+\pi^-$}
\label{fig:dalpc}
\end{minipage}
\end{figure}

The reconstruction in the ``neutral'' mode is based on LKr to construct the u variable and on the spectrometer to define the event charge
and close the kinematics. The possibility to define the u variable using the $\pi^0$'s only is a strong point in this kind of analysis,
because of the charge independence of them. Anyway an alternative u reconstruction can be done using the DCH and the KABES informations, to obtain
a result, useful as cross-check, with different systematic effects. The fiducial region of the detectors is chosen to avoid edge effects. In particular
in order to symmetrize the small difference in the spectrometer acceptance between decays coming from $K^+$ and $K^-$ beams, the spectrometer's 
inner radial cut is chosen according to the actual beam position, periodically measured as the average of the reconstructed transverse vertex position using
three charged pion events. The decay vertex is reconstructed from the $\gamma$ impact point position on the 
LKr, for each $\pi^0$, by using the formula:
\begin{equation}\label{zeta}
Z_{ij}=\frac{1}{m_{\pi^0}}\sqrt{E_i E_j d_{ij}^2}
\end{equation}
Among all the possible $\gamma$ pair, the correct pairing is chosen minimizing the difference between the two $\pi^0$'s vertexes. The
final vertex is obtained as arithmetic average. The kaon invariant mass is obtained including the charged pion measurement from the DCH, in order
to reduce the events background by requiring $|M_{K^\pm}-M_{PDG}|<6MeV/c^2$. A total of $59.3\cdot 10^6$ $K^+$ and $32\cdot 10^6$ $K^-$ 
events has been selected; In plot~\ref{fig:dalpn} the reconstructed Dalitz Plot is shown.
The photon position on the LKr is corrected to take in to account the calorimeter projectivity. The measurement of the charged momentum, slightly
biased by variable DCH misalignment, is corrected exploiting the condition $M_{K^+}=M_{K^-}$ in the $K^{\pm}\rightarrow\pi^{\pm}\pi^+\pi^-$.
Using the same decay mode the magnetic field inversion is monitored online at level of $10^{-3}$ studying the reconstructed kaon mass
with respect to the nominal (PDG) kaon mass. The effect of the residual magnetic field in the decay region (the so called \emph{Blue Field}), mostly 
due to the earth magnetic field, is corrected using the maps obtained from a direct measurement before the data taking.
Several sources of potential systematic bias have been considered. The effect due to the acceptance has been evaluated studying the stability of the 
result varying the cuts definition. The MC has been used to study the contribution of the pion decay in flight to the total systematics error. 
The contribution to the systematics of the online trigger system has been carefully studied. The efficiencies of the L1 and L2, have been evaluated
using data collected with control triggers uncorrelated with the main trigger. In the neutral mode the L1 is essentially obtained with a coincidence between a 
signal (Q1) coming from the scintillating hodoscope and a signal (NTPEAK) compatible with a deposition of four clusters in the LKr. The L2 (MBX) is based
on the algorithm that rejects the $\pi^{\pm}\pi^0$. The main source of systematics comes from the neutral part of the L1 trigger (NTPEAK) being limited by
the number of events in the control sample. In table~\ref{tab:systn} a summary of systematics is shown.
\begin{table}[!t]
\begin{center}
\begin{tabular}{|l|r|}
\hline
\emph{Systematic effect} & \emph{Effect on $\Delta g\cdot 10^{-4}$} \\
\hline
\hline
U calculation \& fitting & $\pm0.2$ \\
LKr non linearity& $\pm0.1$ \\
Shower overlapping & $\pm0.5$ \\
Pion decay & $\pm0.2$ \\
Spectrometer Alignment \& Momentum scale & $<\pm0.1$ \\
Accidentals & $\pm0.2$ \\
L1 Trigger: Q1 & $\pm0.1$ \\
L1 Trigger: NTPEAK & $\pm0.8$\\
L2 Trigger & $\pm0.6$ \\
\hline
Total systematic uncertainty & $\pm1.2$ \\
\hline
\end{tabular}
\end{center}
\caption{Systematics in $K^{\pm}\rightarrow\pi^{\pm}\pi^0\pi^0$}
\label{tab:systn}
\end{table}
The preliminary result in the slope difference using the whole statistics is:
$$
\Delta g=(2.7\pm 2.0_{stat} \pm 1.2_{syst} \pm 0.3_{ext})\times 10^{-4}
$$
where the external error is due to the $\sim3\%$ error~\cite{Eidelman:2004wy}  on the knowledge\footnote{This contribution becomes negligible using the new $g_0$
measurement~\cite{Yao:2006px}} of the g value. 
The resulting charge asymmetry parameter is:
$$
A_g^n=(2.1\pm1.6_{stat}\pm1.0_{syst}\pm0.2_{ext})\times10^{-4}=(2.1\pm1.9)\times 10^{-4}
$$
This result is fully compatible with the SM prediction and is almost one order of magnitude better than the previous measurements~\cite{Smith:1973bi}~\cite{Akopdzhanov:2004xb}. 

\subsection{Results in $K^{\pm}\rightarrow\pi^{\pm}\pi^+\pi^-$}
The offline reconstruction of the $K^{\pm}\rightarrow\pi^{\pm}\pi^+\pi^-$ is totally based on the Spectrometer. The decay vertex is obtained extrapolating
the track segment from the first two chambers from the spectrometer to the decay volume, taking into account the presence of the Blue Field
in the decay region. The track momentum is rescaled, to compensate the variation of the DCH alignment, as described above. In the three charged pion mode 
the chambers' acceptance and the spectrometer performance are most critical with respect to the neutral case where the spectrometer is used only to tag the event.
In particular the beam pipe crossing the chambers determines the main difference in the acceptance between the two beams for which the beam optic can not
control the transverse position better than $\sim 1 mm$.
The cut centered on the actual beam position (at level of the first and last chamber) must be applied to all the pions, resulting in a reduction of $\sim 12 \%$
of the whole statistics. In plot~\ref{fig:dalpc} the reconstructed Dalitz Plot is shown. The only relevant physical background come from the pion decay in flight. More than
$2\cdot10^9$ $K^+$ and $1.1\cdot10^9$ $K^-$ decays are collected in this channel.
The main systematics in the charged mode comes from the pion decay in flight as reported in table~\ref{tab:systc} where the contributions to the systematic error
are summarized.
\begin{table}[!t]
\begin{center}
\begin{tabular}{|l|r|}
\hline
\emph{Systematic effect} & \emph{Effect on $\Delta g\cdot 10^{-4}$} \\
\hline
\hline
Spectrometer alignment & $\pm0.1$ \\
Momentum scale & $\pm0.1$ \\
Acceptance and Geometry & $\pm0.2$ \\
Pion decay & $\pm0.4$ \\
Accidentals & $\pm0.2$ \\
Resolution effects & $\pm 0.3$ \\
L1 Trigger: Q1 & $\pm0.3$ \\
L2 Trigger & $\pm0.3$ \\
\hline
Total systematic uncertainty & $\pm0.7$ \\
\hline
\end{tabular}
\end{center}
\caption{Systematics in $K^{\pm}\rightarrow\pi^{\pm}\pi^+\pi^-$}
\label{tab:systc}
\end{table}
A simpler fitting function can be used in the charged mode case with respect to (\ref{ratio4}) to extract the $\Delta g$ value, exploiting the relative smallness
of the measured g:
$$
R(u)\propto \bar{R}(1+\Delta g u)^4
$$
The preliminary result, based on the 2003+2004 data taking, is:
$$
\Delta g=(0.6\pm 0.7_{stat} \pm 0.7_{syst})\times 10^{-4}
$$
leading to a charge asymmetry parameter of
$$
A_g^c=(-1.3\pm1.5_{stat}\pm1.7_{syst})\times10^{-4}=(-1.3\pm2.3)\times10^{-4}
$$
Also in this case the goal to increase by a factor 10 the sensitivity with respect to the previous measurement has been reached.
The reason for a similar precision of asymmetry results in ``neutral'' and ``charged'' mode, in spite of the different statistics, lies in the fact
that the Dalitz Plot density is most favourable in the ``neutral'' mode.

\section{Dalitz plot parameters measurement in $K^{\pm}\rightarrow\pi^{\pm}\pi^0\pi^0$}
\begin{figure}[!tbh]
\begin{center}
\epsfig{file=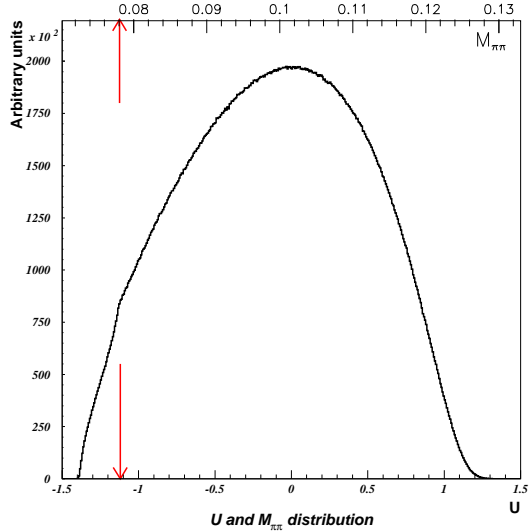,height=3.0in}
\caption{U distribution for $K^{\pm}\rightarrow\pi^{\pm}\pi^0\pi^0$. The $\pi^0\pi^0$ invariant mass square ($M_{00}^2$) is related to the u variable
by $u=(M_{00}^2-s_0)/m_{\pi}^2$. In the arrow position (corresponding to $(2 m_{\pi^+})^2$) it is possible to see the ``cusp'' structure.}
\label{fig:ucusp1}
\end{center}
\end{figure}
In fig.~\ref{fig:ucusp1} the $\pi^0\pi^0$ invariant mass square ($M_{00}^2$) distribution (proportional to the u distribution) is shown. The change of slopes, seen for the 
first time by NA48/2 at $M_{00}^2=(2m_{\pi^+})^2$, can not be explained by the simple matrix element parametrization given in (\ref{eqn:matrixel}). 
This structure has been interpreted by Cabibbo~\cite{Cabibbo:2004gq} as due to the rescattering process $\pi^+\pi^-\rightarrow\pi^0\pi^0$  coming from 
the $K^{\pm}\rightarrow\pi^{\pm}\pi^+\pi^-$ decay.
In fact the $K^{\pm}\rightarrow\pi^{\pm}\pi^0\pi^0$ amplitude can be written as the sum of two terms (just considering the first rescattering order):
the direct $\pi^0\pi^0$ emission $\mathcal{M}_0$, parametrized by the standard polynomial expansion, and the terms due to the 
rescattering process $\mathcal{M}_1$. This last term, proportional to the difference $(a_0-a_2)$ between the pionic scattering length for I=0 and I=2, 
is real below the threshold of $2m_{\pi^+}$ and imaginary above. The total amplitude can be written as:
\begin{equation}\label{eqn:mtotm0m1}
\mathcal{M}^2=
\left\{
\begin{array}{rl}
(\mathcal{M}_0)^2+(\mathcal{M}_1)^2+2\cdot\mathcal{M}_0\mathcal{M}_1 & s_{\pi\pi}<4m_{\pi^+}^2 \\
(\mathcal{M}_0)^2+|\mathcal{M}_1^2| & s_{\pi\pi}>4m_{\pi^+}^2 \\
\end{array}
\right.
\end{equation}
The $\mathcal{M}_0\mathcal{M}_1$ term gives a destructive interference below threshold. Other rescattering diagrams can be 
included in a systematic way as shown by Cabibbo and Isidori~\cite{Cabibbo:2005ez}.
\begin{figure}[!tbh]
\begin{minipage}[t]{0.46\textwidth}
\centering
\includegraphics[width=\textwidth]{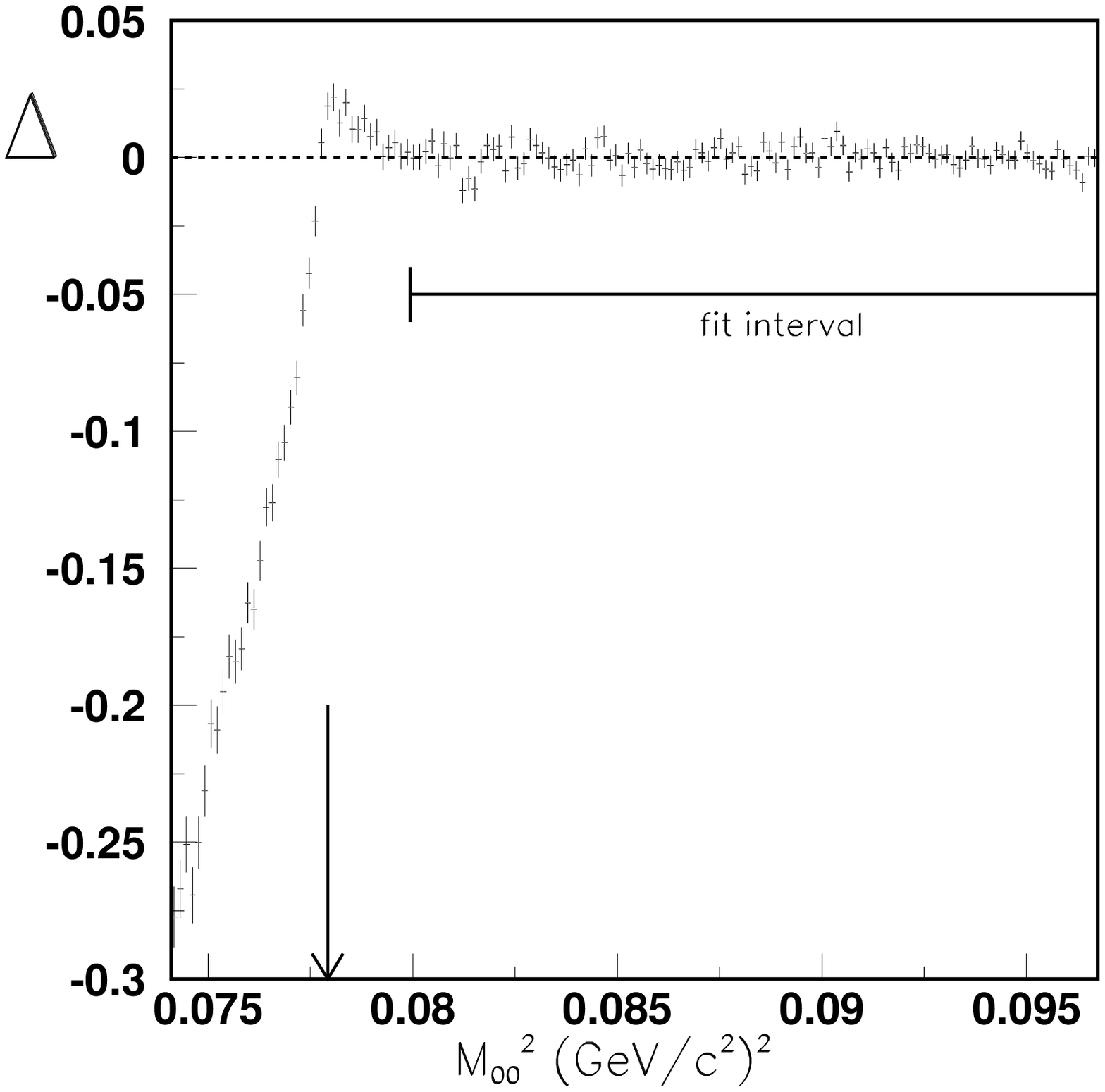}
\caption{$\Delta=(data-fit)/data$ is shown as a function of $M_{\pi^0\pi^0}$. The cusp threshold is indicated by an arrow. 
The fit $\chi^2$ is reasonable only if the fit is restricted to the indicated fit interval.}
\label{fig:fitm0}
\end{minipage}\hfill
\begin{minipage}[t]{0.46\textwidth}
\centering
\includegraphics[width=\textwidth]{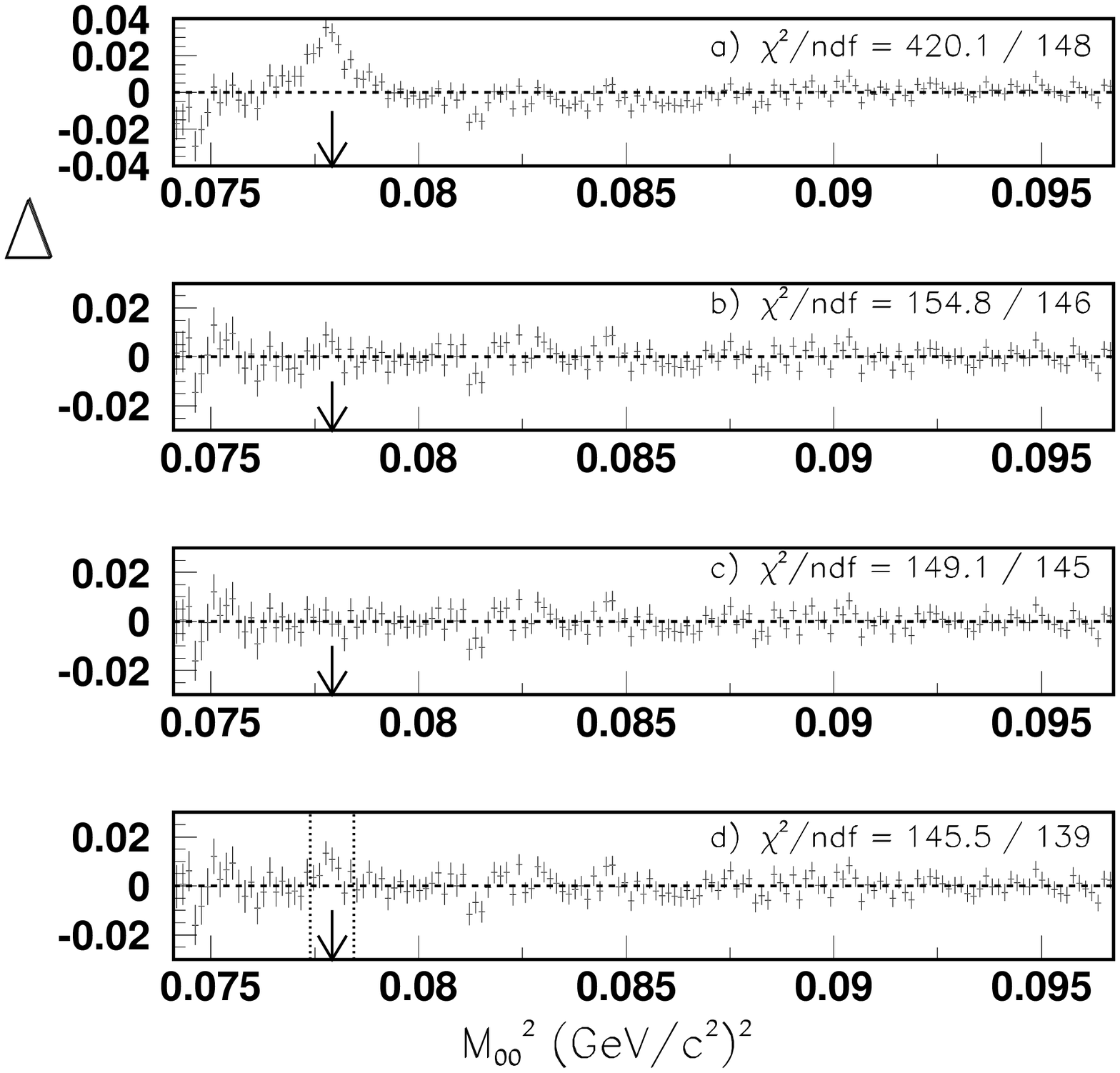}
\caption{The application of the rescattering hypothesis improves the quality of the fit (see text)}
\label{fig:fitm12}
\end{minipage}
\end{figure} 
Fitting data with only the $\mathcal{M_0}$ term yields a fair agreement only above the $2 m_{\pi^+}$ threshold, because of the anomaly introduced 
by the pions strong rescattering in the $M_{00}^2$ distribution, while the standard
expansion is not enough to explain the complex dynamics contributing to the whole decay amplitude.
Using the Cabibbo one-loop model, the fit quality increase giving a $\chi^2$ of 420.1 for 148 degree of freedom, as shown in fig.~\ref{fig:fitm12} .
Including the two-loop Cabibbo-Isidori approach the $\chi^2$ becomes more reasonable (158.8 for 146 degree of freedom). 
Near the threshold the relative pions velocity decreases and the possibility 
to have electromagnetic $\pi^+\pi^-$ bound states increases. However the description of the so called pionic atoms (pionium) 
needs particular care, because of the Coulomb interaction correction and the critical experimental resolution (the fit obtained including
the pionium is shown in the third plot in fig.~\ref{fig:fitm12}).
 For that reason 
we prefer to exclude 7 bins around the threshold position to perform the final fit (last plot in fig.\ref{fig:fitm12} ) in which we have a 
$\chi^2$ of 145.5 for 139 degree of freedom.
The results~\cite{Batley:2005ax}, based on $23\cdot10^6$ events collected in 2003, are obtained setting k, the quadratic v slope, to 0:
$$
g^0=0.645\pm0.004_{stat}\pm0.009_{syst}
$$
$$
h^{'}=-0.047\pm0.012_{stat}\pm0.011_{syst}
$$
where $h^{'}+(1/4)g^2=h$.
The data are compatible with a non zero value for the k parameter, never measured before.
The value of the fit
$$
k=0.0097\pm0.0003_{stat}\pm0.0008_{syst}
$$
is still preliminary. The $(a_0-a_2)$ value is not affected by the $k\neq0$ term, but the g and h values change respectively by $\sim2\%$ and $\sim25\%$.

\section{Dalitz plot parameters measurement in $K^{\pm}\rightarrow\pi^{\pm}\pi^+\pi^-$}
\begin{table}[tb]
\begin{center}
\begin{tabular}{|l|c|c|}
\hline
 & NA48/2 results & PDG06 \\
\hline
g & $(-21.131\pm0.009_{stat}\pm0.012_{syst})\%$ & $(-21.54\pm0.35)\%$ \\ 
h & $(1.829\pm0.015_{stat}\pm0.036_{syst})\%$ & $(1.2\pm0.8)\%$ \\
k & $(-0.467\pm0.005_{stat}\pm0.011_{syst})\%$ & $(-1.01\pm0.34)\%$ \\
\hline
\end{tabular}
\caption{Results for Dalitz Plot parameter in charged mode.}
\label{tab:dalcres}
\end{center}
\end{table}
Thanks to the huge statistics collected in $K^{\pm}\rightarrow\pi^{\pm}\pi^+\pi^-$ decay and to the well tuned MC, very precise measurement of the 
Dalitz plot parameters can be performed. The pion rescattering effects can influence the matrix element also in the ``charged'' mode but, being on the
border of the Dalitz Plot, is not so evident like in the ``neutral'' mode.  
The 
goal of the first and preliminary study presented here is to measure the parameters in the standard polynomial expansion verifying the
validity of (\ref{eqn:matrixel}). The parameters (g,h,k) are obtained minimizing the 
$$
\chi^2(g,h,k,N)=\sum_{u,v}\frac{(F_{data}-NF_{MC})^2}{\delta F^2_{data}+N^2 F^2_{MC}}
$$
where F represents the population (in data or in MC) in the (u,v) bin. The MC population is obtained adding the 4 components corresponding to
the four possible terms in (\ref{eqn:matrixel}). The relative weights, obtained from the fit, are the polynomial expansion parameters.
The coulomb correction is applied to take into account the pion electromagnetic interaction.
The main contributions to the systematic uncertainty, at this analysis stage, come from the momentum scale in the charged pions measurement, 
the kaon momentum spectrum in the MC and trigger inefficiencies. 
The preliminary result is based on $0.47\cdot10^9$ events collected in the 2003 run. In table~\ref{tab:dalcres} the results are presented
and the agreement with the PDG06 values is shown. The previous measurements, by experiments made in 70s, are one order of magnitude less precise 
with respect to the NA48/2 measurement, based on $~1/4$ of the whole statistics.
 
\section{Conclusions}
The main goal of the NA48/2 experiment was to measure, with a precision at level of $10^{-4}$ the charge asymmetry parameter $A_g$, 
both in $K^{\pm}\rightarrow\pi^{\pm}\pi^+\pi^-$ and in  $K^{\pm}\rightarrow\pi^{\pm}\pi^0\pi^0$ decays. 
The preliminary results obtained in the ``neutral'' ($A_g^n$) and ``charged'' ($A_g^c$) mode:
$$
A_g^n=(2.1\pm1.6_{stat}\pm1.0_{syst}\pm0.2_{ext})\times10^{-4}=(2.1\pm1.9)\times 10^{-4}
$$
$$
A_g^c=(-1.3\pm1.5_{stat}\pm1.7_{syst})\times10^{-4}=(-1.3\pm2.3)\times10^{-4}
$$
are compatible  with the SM predictions and with our previous results based on partial samples~\cite{Batley:2006neu}~\cite{Batley:2006mu}.

In $K^{\pm}\rightarrow\pi^{\pm}\pi^0\pi^0$ the standard polynomial matrix element expansion is not enough to describe the observed
$\pi^0\pi^0$ invariant mass spectrum. Taking into account the rescattering processes, whose contributions are proportional to
the $(a_0-a_2)$ term the corresponding slope are found to be (setting k=0)
$$
g^0=0.645\pm0.004_{stat}\pm0.009_{syst} \qquad h^{'}=-0.047\pm0.012_{stat}\pm0.011_{syst}
$$
However a $\sim1\%$  for the k quadratic slope is obtained with a complete fit (preliminary result).

In the charged mode the standard polynomial fit has been employed. The Dalitz plot parameters have been remeasured with 
higher precision with respect to the previous old measurement.

\end{document}